\documentclass[11pt]{article}
\usepackage[utf8]{inputenc}
\usepackage{amsmath}
\usepackage{amssymb}
\usepackage{booktabs}
\usepackage{fullpage}
\usepackage{graphicx}
\usepackage{verbatim}
\usepackage{url}
\usepackage{caption}
\usepackage{subcaption}
\usepackage{booktabs}
\usepackage{floatrow}
\usepackage{color}
\usepackage{enumitem}
\usepackage{multirow}
\usepackage{authblk}

\title{How does the Topology of City Streets Impact on their \\ Respective
Optimization?}
\author[1]{Eric K. Tokuda}
\author[2]{Cesar H. Comin}
\author[1]{Luciano da F. Costa}
\affil[1]{University of São Paulo, São Carlos, Brazil}
\affil[2]{Federal University of São Carlos, São Carlos, Brazil}
\date{}

\begin{document}
\maketitle

\abstract{
Several natural and artificial structures and systems are somehow optimized for performing
specific functionalities.  The structure and topology of cities is no exception, as it
is critically important to ensure effective access to the several resources as well as
overall mobility.  The present work addresses the important subject of improving the
plan of a given city through the incorporation of avenues and other express ways
such as bridges and tunnels.  More specifically, we start with the topology of a real
city and consider the incorporation of a express way between any two locations in the city, keeping one location fixed and varying the angle of the other.  The whole city area is covered in this manner, which allows us to derive a respective energy surface indicating the gain obtained
regarding the average shortest path length for each of the possible situations.  These surfaces
therefore provide a complete picture of how  much each city can be improved regarding minimal distances.  Quite distinct surfaces have been obtained for 18 considered European cities.  These surfaces are then characterized in terms of the number of local extrema and respective spatial complexity, expressed in terms of a raggedness measurement.  Measurements are also obtained respectively to the geometry and topology of the considered cities.  It is shown that the shortest path gain depends strongly on some of the considered measurements, especially lacunarity and transitivity.  Interestingly, the intricacy of the energy surfaces resulted in relatively little correlation with the topological and geometrical measurements.}

\section{Introduction}

Much of the physical world structures and dynamics can be understood from the perspective of optimization.  Examples
of this important tendency can be observed everywhere.
The water in our glasses remains at its lowest energy
(despite minute stochastic fluctuations), the rivers flow
towards the ocean, and planets and comets follow the smallest
energy orbit (e.g.~\cite{hildebrandt1996parsimonious,sewell1987maximum}).
Indeed, it is very likely that the whole of the physical actions
follow the principle of minimum energy, as described by
variational calculus (e.g.~\cite{troutman2012variational}) .

At the human scale, several related activities also closely
follow optimization principles, especially because of the
often substantial involved expenses and resources.  Examples
of human activity that relate critically to optimization 
include but are by no means limited to the development of
effective housing, telecommunication and highway networks, 
supply chains, civil architecture, and urban planning.
The present work addresses the latter of these.

Because cities (still) provide most of the basic infra-structure
required for daily human needs, great effort has been invested
in optimizing several of its characteristics.  One aspect of
particular relevance concerns the average distance between
different parts of a city (e.g.~\cite{zhan1998shortest}). This objective, however, is typically made more complex by the several additional constraints that are usually imposed, such as the need to avoid big mountains, crossing of rivers, the presence of the coastline, the need of parks, as well as
preservation of nature  and historic patrimony, among many others.

As a consequence of the above tension between the need to optimize
city streets networks and several important constraints, the overall
structure of street networks often result in intricate geometrical
and topological effects, including bottlenecks, borders, etc.

With some exceptions, such as the rebuilt of Paris in 1860~\cite{pinkney2019napoleon} as well as planned
cities, streets networks are modified progressively through the
incremental incorporation of avenues, bridges, tunnels, etc.
Almost invariably, at least at the outset, these resources are
aimed at fast transit flow, therefore constituting express venues
that will hopefully optimize the movements of automobiles.

Given the typically intricate structure of street networks, it
becomes a rather challenging problem to identify which points
of the existing cities could be interconnected in a more express
manner so as to improve the overall transit.  This problem constitutes
the main focus of interest on the present work.

One important related class of problems refers to the network optimization, which is often motivated by real-world applications, such as manufacturing and transportation~\cite{ahuja1995applications}. Different optimization criteria can be used in the optimization, such as flow, matching, and the path length.

A related problem concerns the search for shortcut edges that result in the highest reduction of average path lengths. Several works have already explored variations of this problem and efficient approximation methods have been proposed~\cite{meyerson2009minimizing,demaine2010minimizing,rijkersaugmenting}. Related problems, such as the minimization of the \emph{diameter} of the graph, as opposed to the average path length, have also been extensively explored ~\cite{frati2015augmenting,luo2008computing}.

In~\cite{buhl2006topological}, the authors analyzed the topology of several cities considering the path lengths and the robustness to disconnections. They observed different patterns, such as tree-like and meshed urban environments, which potentially contribute to varying vulnerability to random failures.

A distinguishing feature of the present approach is the systematic
mapping of the effect of all possible considered modifications on the
respective improvement as quantified by the distribution of the average
minimal distances.  The implemented incremental modifications consists 
of adding one express way at a time with length corresponding to 25\% of the diameter of each city.  

Eighteen European cities were considered. The choice was based on the physical areas of the cities, which was chosen to be approximately  $110 km^2$. 
For each city, the addition of one express way was performed 
considering 12 distinct, equally spaced angles. By doing so, four-dimensional maps of the gain in average minimal length were obtained for
each of these possible modifications.  Therefore, a systematic and complete picture of the possible benefits of incorporating an express way could be obtained.  The results differed substantially in distribution of gain and
local maxima, suggesting that the improvements obtained for the considered cities can be quite different as a consequence of their respective
topological and geometrical properties. In order to address this interesting point, several topological and geometrical measurements of the cities were obtained and analysed independently and with respect to the distance gains.

We considered different types of features to characterize the city, describing the topology and the geometry presented by the street network. More specifically, we extracted statistics from the following geometrical measurements: the position of the vertices; the orientation of the streets; and the characterization of holes in the city. The following topological features were also adopted as the means for characterizing the topology of the considered cities: vertex degree and vertex accessibility.

\section{Material and Methods}
\label{sec:method}

To study the effect that the graph structure may have on the 
reduction of the city average path length as new express ways are incorporated into the city, a set of topological and geometrical measurements was extracted from 18 cities. In the following, we describe the data used in the analysis as well as the considered measurements.

\subsection{The dataset}
We consider a graph representation of a city $G(V, E)$, with edges ($e \in E$) representing the segments of streets and vertices ($v \in V$) representing the intersections between streets or dead ends~\cite{barthelemy2011spatial}. 
We considered the Openstreetmap~\footnote{https://www.openstreetmap.org/} data for the extraction of the city graphs. To avoid spatial normalization issues, only cities with similar areas were taken into account. Eighteen (18) European cities with areas between 105 and 115 $km^2$ were therefore chosen, being shown in Table~\ref{tab:cities}.

\begin{table}[]
   \caption{List of the analyzed cities, sorted by area, in square kilometers.}
   \centering
   \begin{tabular}{llc}
   \toprule
      City & Country & Area (km$^2$)\\
   \midrule
Paris&France&105 \\
Gelsenkirchen&Germany&105 \\
Koblenz&Germany&105 \\
Livorno&Italy&105 \\
Kassel&Germany&107 \\
Wilhelmshaven&Germany&107 \\
Estarreja&Portugal&108 \\
Vila Nova de Cerveira&Portugal&108 \\
Worms&Germany&109 \\
Heidelberg&Germany&109 \\
Bristol&UK&110 \\
Gap Hautes-Alpes&France&110 \\
Liverpool&UK&112 \\
Emden&Germany&112 \\
Santa Comba Dão&Portugal&112 \\
Istres&France&114 \\
Jena&Germany&115 \\
Castelo de Paiva&Portugal&115 \\
   \bottomrule
   \label{tab:cities}
   \end{tabular}
\end{table}

\subsection{Graph characterization}
\label{sec:measurements}

A graph can be characterized in several different ways~\cite{costa2007characterization}. For the purpose of the present work, we consider measurements related to the local connectivity, the spatial distribution of the vertices, the distribution of the edges angles, the uniformity of the blocks in the city, and the dynamics of a random walk over the graph. We explain each of these measurements below.

The local connectivity of a graph is commonly studied in terms of the vertex degrees, which is defined as the number of connections that a node makes to other vertices in the graph. The distribution of this property in a given graph can then be characterized in terms of statistics such as the mean and standard deviation.

The transitivity (or clustering coefficient) can complement the degree measurement, corresponding to the tendency of nodes to form clusters, therefore being related to the presence of closed paths~\cite{newman2001scientific}. In an undirected graph, the local transitivity of a vertex quantifies how close its neighbours are to forming a clique. 

The geometry of the blocks in the city was quantified considering the \emph{lacunarity}~\cite{allain1991characterizing,rodrigues2005self}. This concept is often attributed to Mandelbrot, being aimed at complementing the characterization of fractal structures for digital images. One of the main algorithms used for its computation relies on calculating the coefficient of variation of the pixels inside sliding windows. Considering that this algorithm was defined for images, the graph first is embedded into an Euclidean space and rasterized to obtain an image. 

The accessibility of a vertex quantifies its influence over successive neighbourhoods, also reflecting an adopted specific dynamics that is of particular importance~\cite{travenccolo2008accessibility}. This measurement has been often used in conjunction with random walks. In this case, given a parameter $h$ corresponding to the order of the neighborhood of the reference node, the accessibility estimates the number of neighbors that can be reached after exactly $h$ steps of the walk. We quantify this measurement per vertex and calculate the mean and standard deviation of all values.

\subsection{Express way}
\begin{figure}[]
    \begin{subfigure}[b]{0.2\textwidth}
   \includegraphics[width=\textwidth]{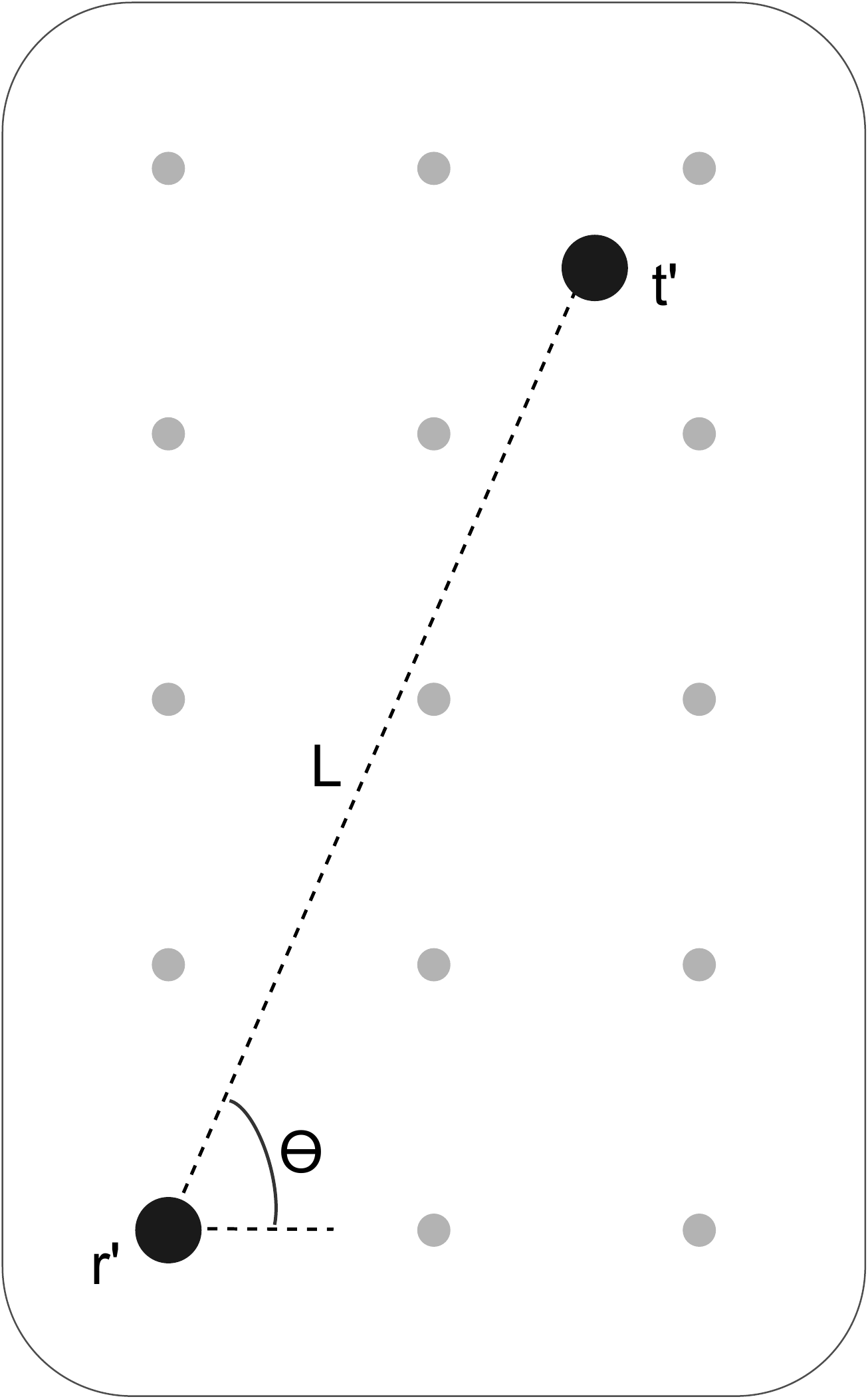}
       \centering
   \caption{Grid}
    \end{subfigure} \quad
    \begin{subfigure}[b]{0.2\textwidth}
    \centering
   \includegraphics[width=\textwidth]{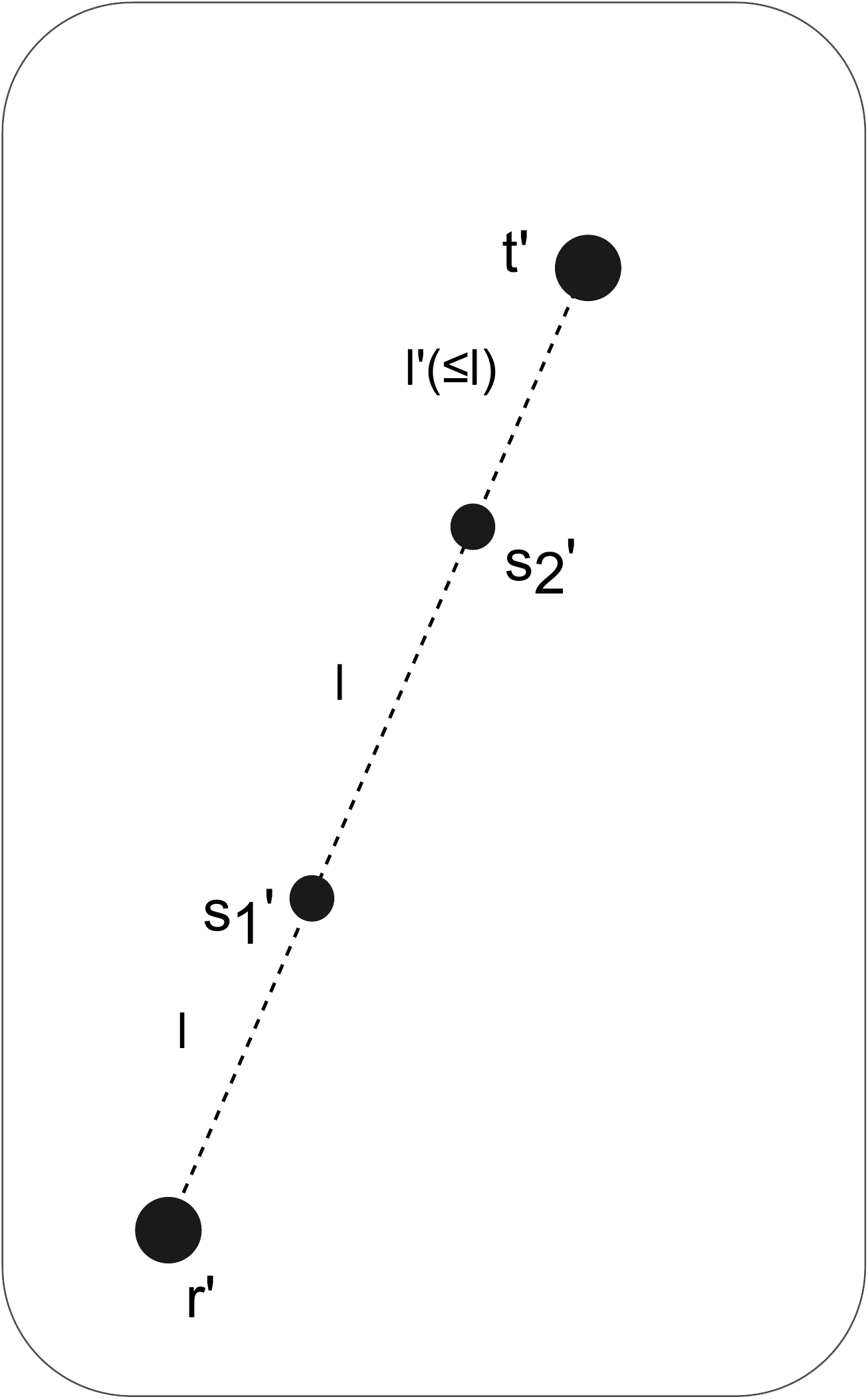}
   \caption{Reference points}
    \end{subfigure} \quad
    \begin{subfigure}[b]{0.2\textwidth}
    \centering
   \includegraphics[width=\textwidth]{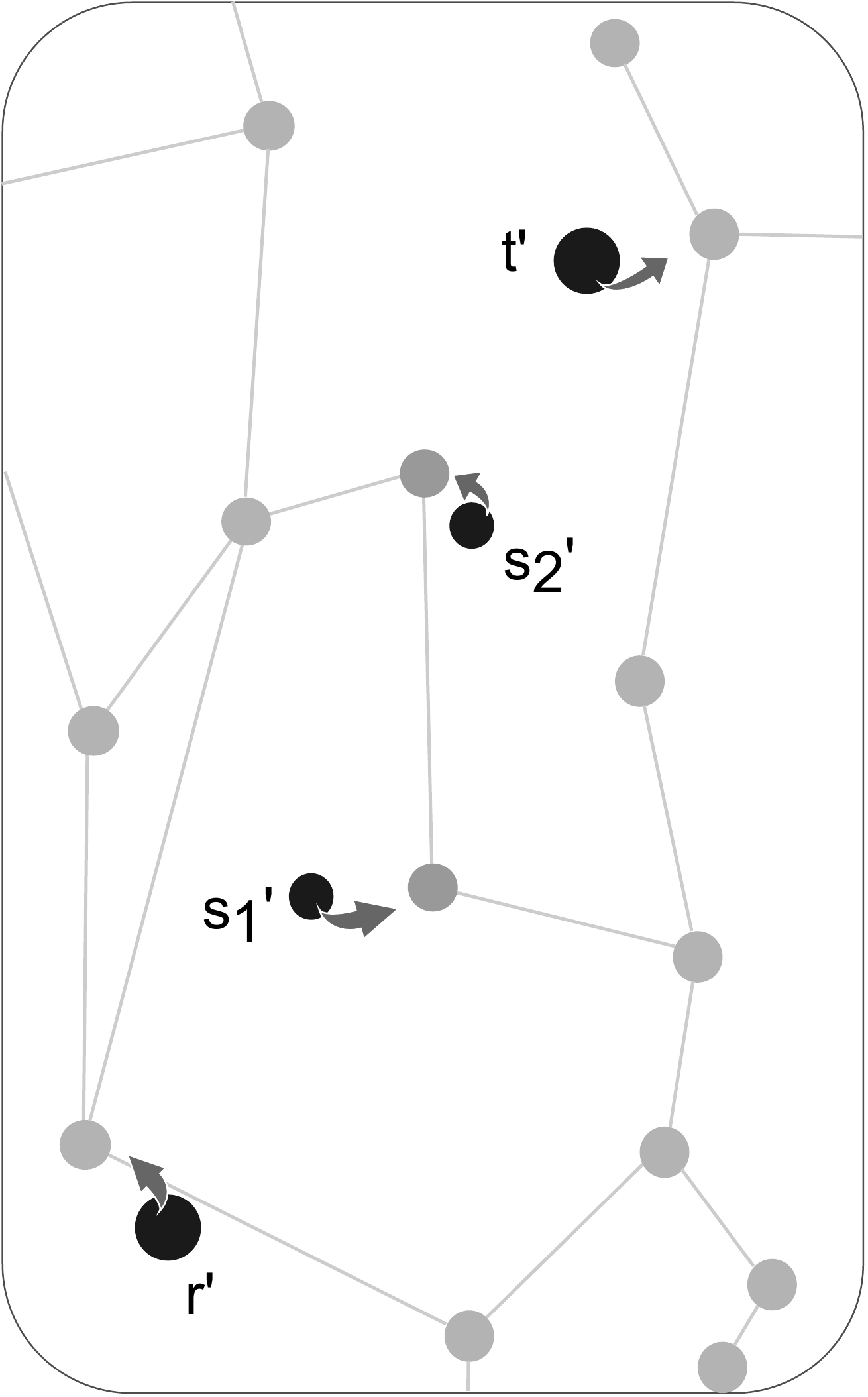}
   \caption{Closest vertices}
    \end{subfigure} \quad
    \begin{subfigure}[b]{0.2\textwidth}
    \centering
   \includegraphics[width=\textwidth]{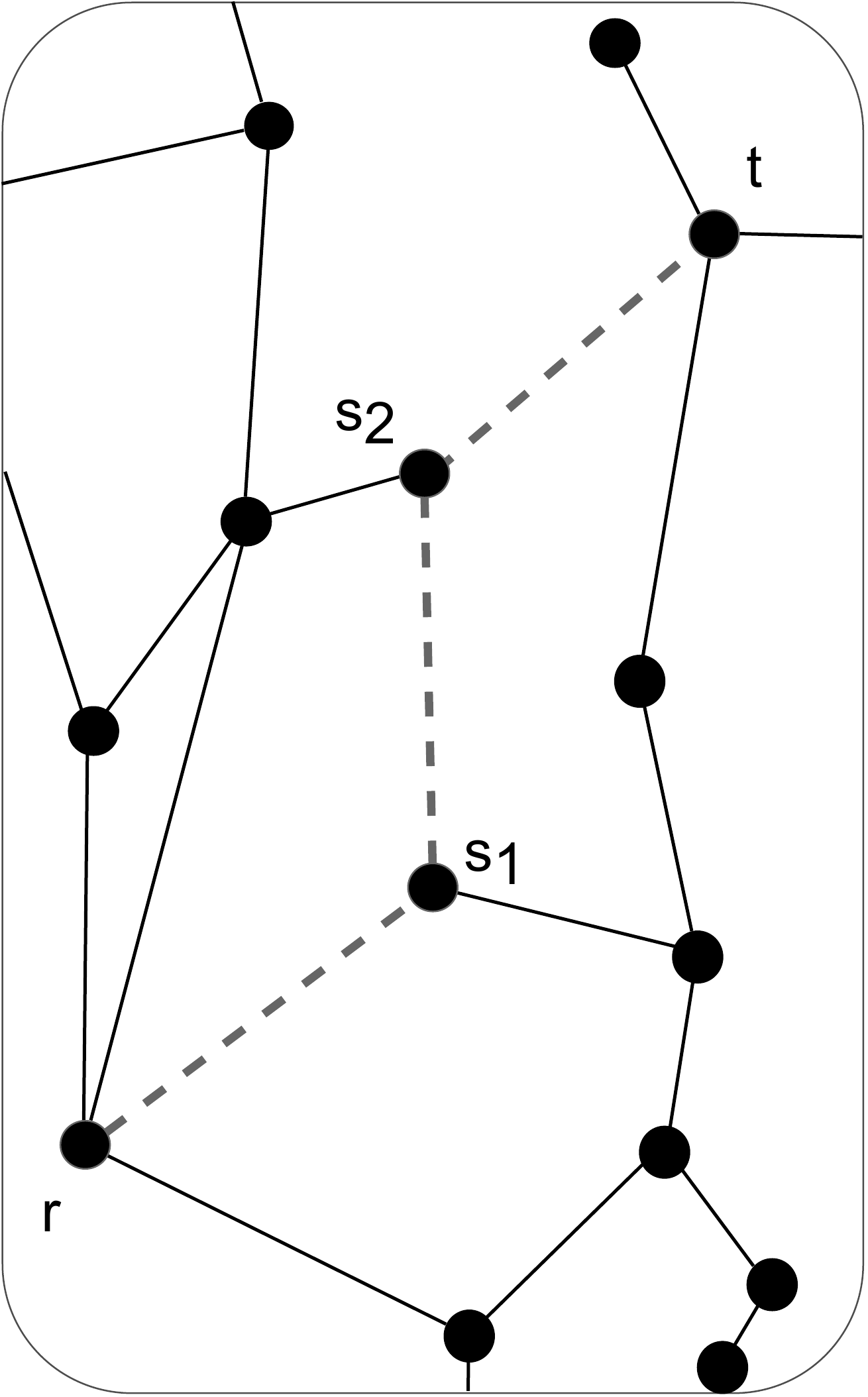}
   \caption{Avenue}
    \end{subfigure}
    \caption{Addition of an express way to the graph. (a) An evenly-spaced set of points over the graph region is sampled. Given one point of the set $r'$, a point $t'$ is determined considering a direction $\Theta$ and length $L$. (b) The segment $\overline{r't'}$ is partitioned by the intermediate points, $s_1'$ and $s_2'$, such the inner segments have same length, $d$, except eventually by one segment, which may be lesser than l. (c) The closest vertices in the graph, $r,s_1,s_2,t$, to the vertices $r',s_1',s_2',t'$, respectively, are determined. (d) The express way is defined as the path $(r,s_1,s_2,t)$. The edges $rs_1$ and $s_2t$ did not exist and were added to the graph.}
    \label{fig:avdiagram}
\end{figure}

In this work, an \emph{express way} is composed of edges $\overline{v_1v_2}$, $\overline{v_2v_3}\ldots\overline{v_{m-1}v_m}$ and is represented as $(v_1,v_2,\ldots,v_m)$. Figure~\ref{fig:avdiagram} illustrates the construction of each express way. The procedure works as follows:

\emph{(a) Grid over the graph.} First, a grid defining a set of points $R$ is positioned on top of the city. Also, an evenly spaced set of angles $\Theta$ is defined. For each combination $(r', \theta) \in P\times\Theta$, an express way will be added.

\emph{(b) Reference points.} Given $r'$ and $\theta$, the segment of line defined by $r'$ and angle $\theta$ is obtained and the point $t'$ on the line, $L$ distant to $r'$ is determined. The segment $\overline{r't'}$ is partitioned into segments of length $l'<= l$, $s_1',\ldots,s_m$, such that the segments $\overline{r's_1},\overline{s_2s_3}\ldots\overline{s_{m-1}s_m}$ have length $l$ and  $\overline{s_mt}$ have length lesser or equal than $l$.

\emph{(c) Closest vertices.} The vertices $r,s_1,\ldots,s_m,t$ are  defined as the vertices of the graph closest to $r',s_1',\ldots,s_m',t'$, respectively.

\emph{(d) Missing edges.} The resulting express way is defined as $(r,s_1,\ldots,s_m,t)$; any non-existent edge is added.

The new graph $G' (V', E')$ created in this way has the same set of vertices ($V=V'$) and with $0\leq|E'|\leq |E|+m+1$. It is important to notice that express ways added in highly connected graphs will likely lead to fewer number of new edges. The systematic addition of express ways is parameterized by the spatial sampling (grid), the angles, the expected express way endpoints length and the spacing between the intermediate points.

\subsection{Gain}
For each new graph $G'$, the euclidean shortest path between all pairs of vertices are calculated and the average path length $a_{G'}$, is calculated. The gain resulting from the added express way is defined as the difference between the average path length of the new graph $G'$ and the original one, $G$. The gain defined this way is always non-negative.
\begin{equation}
gain(G')=a_G - a_{G'}.
\end{equation}

For each city, the gain of all express ways are averaged and pairwise compared to the city graph features through the Pearson correlation.

\subsection{Gain analysis}
\label{subsec:gainanalysis}

The average shortest path length gains calculated for different positions $r'=(x,y)$ on the grid and angles $\theta$ can be represented as a 3D image. In order to do so, we define a 3D image $I$ with size $N_r\times N_c \times N_z$ in which the first two coordinates are related to the grid points and the third coordinate is associated to the angles. Each voxel in the image has size $l\times l\times \delta\theta$, where $l$ is the spacing of the grid and $\delta\theta$ is the angle interval used for generating the express ways. The 3D image is created by associating the gain calculated for each combination $(r',\theta)$ to the respective voxel in the image. 

The generated gain image contains important information regarding the gain values for the cities. Thus, we characterize the images using two properties:  (1) the number of gain peaks and (2) the raggedness of the gain distribution. The number of gain peaks is related to the number of $(r',\theta)$ combinations where the gain can be made locally optimal.  The intricate gain surfaces typically obtained for the considered cities makes the identification of the peaks a particularly challenging issue hardly to be met by using traditional linear filtering approaches such as those based on convolution.  In the present work we used the coincidence similarity index  \cite{costa2021further,costa2021comparing} for that finality. Despite the computational simplicity of this filter, which combines the Jaccard and interiority (or overlap) similarity indices, it actually implements a non-linear action allowing the small scale noise and detail to be substantially attenuated while the large scale peaks are made sharper and narrower~\cite{costa2021comparing}. After enhancing the peaks by using the aforementioned methodology, they are identified by finding local maxima in the 3D image. Peaks are defined as voxels having value larger than their neighbors. In order to exclude background peaks, we only keep peaks having value larger than the median of the image.

In this work, the \emph{raggedness} is regarded as a quantification of the complexity of the obtained gain distribution, which on its turn reflects the changes of gain obtained as the position and angle of the express ways is altered. This measurement is calculated as follows. First, a threshold $t$ is applied to the 3D gain image. Thus, values larger than or equal to $t$ become 1 and values smaller than $t$ become 0. The result is a binary image in which voxels having value 1 correspond to large gains. Such voxels define regions in the $R\times\Theta$ space where the addition of express ways tends to yield large gains. Next, the following quantity is calculated:

\begin{equation}
    S = \frac{\pi^{1/3}(6V)^{2/3}}{A}.
\end{equation}

Where $V$ is the volume of the aforementioned regions in the $R\times\Theta$ space and $A$ is the surface area of these regions. $S$ is called the \emph{sphericity} of the gain since it quantifies the similarity of the voxels in the binary gain image to a sphere~\cite{da2010shape}. If $S=1$, it means that the variation of gain values is relatively simple, while small values of $S$ indicate that the gain varies in a complex manner according to the position and angles of the express ways. The raggedness is then defined as

\begin{equation}
    H = \frac{1}{S}.
\end{equation}

\section{Experiments and discussion}

\begin{figure}[]
    \begin{subfigure}[b]{0.7\textwidth}
   \includegraphics[width=\textwidth]{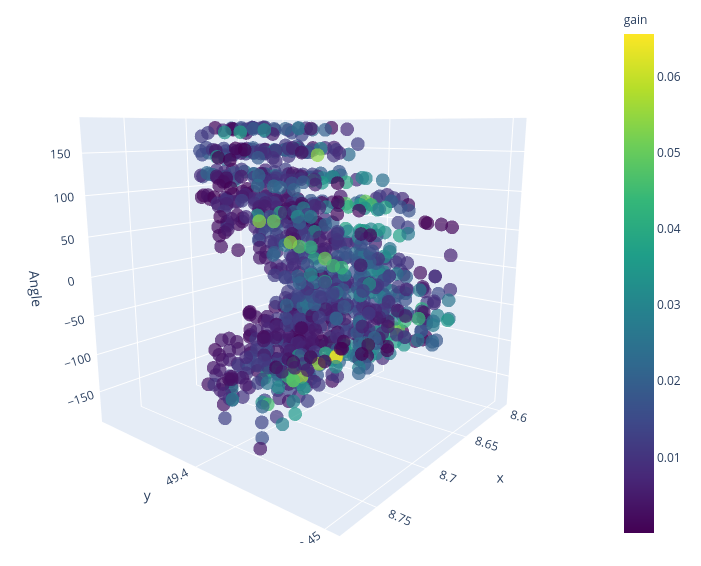}
       \centering
   %\caption{Heidelberg}
   \label{fig:caption1}
    \end{subfigure} \quad
    \caption{3-D visualization of the gains of the proposed approach considering the city of Heidelberg, Germany. The axes $x,y,gain$ correspond to the spatial coordinates (x and y), and to the average gain over all express ways, respectively.}
    \label{fig:vis3d}
\end{figure}

Initially, the network loops and multiple edges were removed. Also, only the largest connected component was considered. Following Section~\ref{sec:method}, the parameters considered in the experiments were $L=0.25 d$, where $d$ is the diameter of the city.  This is believed to correspond to a reasonable extension of an express way, while reflecting the likely situation that the demands and expenses scales proportionally with the size of the cities.  The spacing between intermediate points was fixed at $l=500m$, while the angles were evenly spaced by $30^\circ$ for computational reasons. Figure~\ref{fig:vis3d} shows a visualization of the gains obtained for one of the cities.

Two categories of graph measurements have been considered, namely derived from the graph topology and from the graph geometry. The considered topological measurements are based on the vertex degree distribution and the local vertex transitivity. From the vertex degree, the mean and the standard deviation are calculated. Also, the fraction of vertices with degrees 3, 4, and 5 are calculated. The dynamics occurring on the graphs are studied considering the accessibility of the vertices. The accessibility for the random walk dynamics is characterized through the mean and the standard deviation.

The geometry of the city is studied through the angling of the edges, the spatial distribution of the vertices and the lacunarity. To account for the distribution of edge orientations and vertex positions, the standard deviation is calculated for each measurement.  One central parameter of the calculus of the lacunarity is the radius, which in this work has been considered as 210 meters.

\subsection{Topological and geometrical analysis}

The topological and geometrical features described in Section~\ref{sec:measurements} were calculated for each city in Table~\ref{tab:cities}. Table~\ref{tab:featdescription} contains a description of each measurement. The values obtained for the topological measurements are shown in Table~\ref{tab:valuestop}. 

\begin{table}[]
   \caption{Summary of the features considered in the graph characterization.}
   \centering
   \begin{tabular}{lll}
   \toprule
	   Notation &Category& Feature description \\ \midrule
	   acavg &topological& Average of the vertex accessibility  \\
	   acstd &topological& Standard deviation of the vertex accessibility  \\
	   deavg &topological& Average of the vertex degrees  \\
	   destd &topological& Standard deviation of the vertex degrees  \\
	   degr3 &topological& Fraction of vertices with degree=3  \\
	   degr4 &topological& Fraction of vertices with degree=4  \\
	   degr5 &topological& Fraction of vertices with degree=5  \\
	   travg &topological& Average of the vertex local transitivity \\
	   trstd &topological& Standard deviation of the vertex local transitivity  \\
	   lacun &geometrical& Lacunarity \\
	   eostd &geometrical& Standard deviation of the edge orientations \\
	   vpstd &geometrical& Standard deviation of the vertex positions  \\
   \bottomrule
   \label{tab:featdescription}
   \end{tabular}
\end{table}

\begin{table}[]
   \caption{Topological features of the cities. The lowest values are emphasized as italicized and the greatest as bold.}
   \footnotesize
\centering
\begin{tabular}{lrrrrrrrrr}
\toprule
City &  deavg &  destd &  degr3 &  degr4 &  degr5 &  travg &  trstd &  acavg &  acstd \\
\midrule
Bristol &    4.546 &   1.826 & 0.072 & 0.092 & 0.046 &      0.058 &     0.128 &     16.618 &     9.058 \\
Castelo de Paiva &    4.468 &   1.791 & 0.112 & 0.096 & 0.049 &      0.111 &     0.161 &     15.895 &     \textit{7.155} \\
Emden &    4.749 &   1.875 & \textit{0.032} & \textit{0.059} & 0.059 &      0.056 &     0.129 &     15.637 &     8.293 \\
Estarreja &    4.796 &   1.834 & 0.090 & 0.082 & 0.038 &      0.087 &     0.150 &     19.203 &     8.797 \\
Gaphautes-alpes &    4.100 &   1.698 & 0.175 & 0.164 & 0.052 &      \textbf{0.122} &     0.165 &     \textit{14.637} &     8.101 \\
Gelsenkirchen &    4.780 &   1.810 & 0.077 & 0.161 & 0.037 &      0.093 &     0.155 &     24.083 &    \textbf{12.521} \\
Heidelberg &    4.648 &   1.695 & 0.153 & 0.156 & 0.124 &      0.088 &     0.156 &     23.809 &    11.483 \\
Istres &    4.095 &   1.754 & 0.180 & 0.142 & 0.037 &      0.120 &     \textbf{0.175} &     14.643 &     7.455 \\
Jena &    4.829 &   1.866 & 0.053 & 0.089 & 0.068 &      0.075 &     0.153 &     18.939 &     9.755 \\
Kassel &    4.888 &   1.814 & 0.078 & 0.117 & 0.069 &      0.071 &     0.144 &     24.383 &    12.233 \\
Koblenz &    4.377 &   1.767 & 0.167 & 0.113 & 0.080 &      0.085 &     0.150 &     18.716 &     9.825 \\
Liverpool &    4.558 &   1.899 & 0.051 & 0.090 & 0.038 &      \textit{0.044} &     \textit{0.114} &     18.387 &    11.982 \\
Livorno &    4.277 &   1.616 & 0.212 & 0.181 & 0.126 &      0.107 &     0.167 &     20.298 &     9.406 \\
Paris &    \textit{3.863} &   \textit{1.257} & \textbf{0.408} & \textbf{0.253} & \textbf{0.152} &      0.071 &     0.137 &     \textbf{34.264} &    10.772 \\
Santa Comba Dão &    4.638 &   1.774 & 0.129 & 0.120 & \textit{0.031} &      0.116 &     0.166 &     17.744 &     7.882 \\
Vila Nova de Cerveira &    4.909 &   1.682 & 0.112 & 0.098 & 0.034 &      0.114 &     0.169 &     20.164 &     7.987 \\
Wilhelmshaven &    4.764 &   \textbf{1.932} & 0.063 & 0.089 & 0.081 &      0.067 &     0.142 &     20.868 &    12.106 \\
Worms &    \textbf{4.976} &   1.752 & 0.078 & 0.099 & 0.072 &      0.075 &     0.150 &     20.856 &     8.687 \\
\bottomrule
\end{tabular}

   \label{tab:valuestop}
\end{table}

A similar average degree was obtained for all cities, except for Paris which presents
a slightly smaller value.  Similar standard deviations of the node degree, as well as the average local transitivity, can also be observed throughout.  
Interestingly, the number of nodes with degree 3, 4 or 5  varied more substantially.

Therefore, we can conclude that all the 18 considered cities present similar topologies as far as most of the adopted measurements are concerned.  The observed dispersion among the cities for degr3, degr4 and degr5 reveals a local heterogeneity of connections, which can be directly
associated to the presence of street dead ends and irregularly shaped blocks, with the
variations of degr5 being especially related to the presence of intersections with distinct number of arms.

Large dispersions can be observed for all the three adopted accessibility measurements.
The highest accessibility is observed for Paris, which is a direct consequence of the higher uniformity of this city.  The smallest average accessibility was observed for
Gaphautes-Alpes, with Istres presenting a very similar value.
Further substantiating the high topological heterogeneity already revealed by the average accessibility, the respective standard deviation also results markedly distinct. 

In summary, the 18 networks in our analysis present substantial overall similarity of measurements related to the vertex degree, while showing divergence in the measurements related to the local transitivity and the accessibility.  

The geometrical features are shown in Table~\ref{tab:valuesgeom}. Relatively little variation can be observed for the edge orientation standard deviation, indicating
respective geometric uniformity.  However, larger variations characterize the other two adopted measurements.  In the case of the lacunarity, it reaches its minimum value for
Paris, reflecting the geometric uniformity of this city, while the higher intricacy of the port city of Livorno implied a substantially higher lacunarity value.  The other cities
present relatively similar lacunarity.  The vertex position standard deviation also resulted markedly distinct among the analyzed cities, reaching its maximum for Livorno, and its minimum for
Istres.  

\begin{table}[]
   \caption{Geometric features of the cities. The lowest values are emphasized as italicized and the greatest as bold.}
   \centering
\begin{tabular}{lrrr}
\toprule
City &  lacun &  vpstd &  eostd \\
\midrule
Bristol &    1.100 &     0.269 &    0.761 \\
Castelo de Paiva &    1.236 &     0.283 &    0.822 \\
Emden &    1.287 &     0.216 &    0.831 \\
Estarreja &    1.177 &     0.297 &    0.823 \\
Gaphautes-alpes &    1.266 &     \textit{0.193} &    0.842 \\
Gelsenkirchen &    1.135 &     0.309 &    0.785 \\
Heidelberg &    1.196 &     0.236 &    \textbf{0.872} \\
Istres &    1.280 &     0.222 &    0.866 \\
Jena &    1.208 &     0.250 &    0.791 \\
Kassel &    1.121 &     0.278 &    0.801 \\
Koblenz &    1.210 &     0.209 &    0.807 \\
Liverpool &    1.097 &     0.290 &    \textit{0.755} \\
Livorno &    \textbf{1.400} &     \textbf{0.345} &    0.820 \\
Paris &    \textit{1.070} &     0.310 &    0.769 \\
Santa Comba Dão &    1.275 &     0.284 &    0.844 \\
Vila Nova de Cerveira &    1.282 &     0.293 &    0.859 \\
Wilhelmshaven &    1.253 &     0.236 &    0.807 \\
Worms &    1.189 &     0.257 &    0.832 \\
\bottomrule
\end{tabular}

   \label{tab:valuesgeom}
\end{table}

Though smaller, \texttt{vpstd} variations are observed also for the other
18 cities.  In other words, the considered cities can be understood as being relatively
similar regarding \texttt{eostd} and lacunarity, but present a more accentuated dispersion of positional standard deviation.  This reveals that, though with similar shapes (directly related to the number of adjacent streets), the block sizes vary more substantially between the considered cities.  As a consequence, the local topology of the city would not change substantially. However, the variation of block sizes can be expected to have
implications on the obtained average path length gain. For instance, if a square block is too long, it will imply less crossings along its extension, therefore influencing the topology at a larger scale.

\begin{figure}[]
   \centering
   \includegraphics[width=0.9\textwidth]{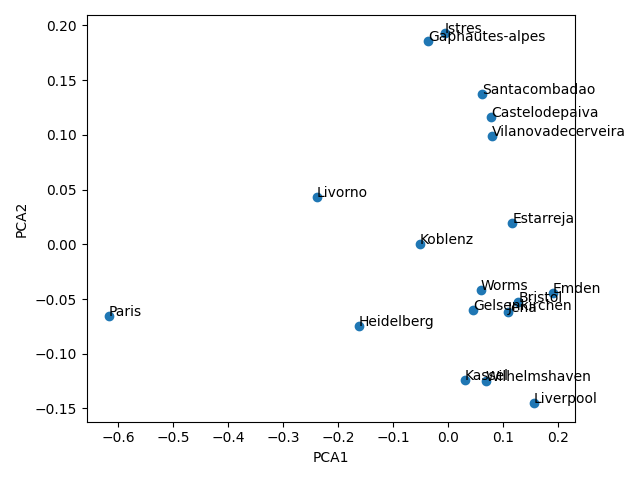}
   \caption{Principal component analysis of the features of the cities. The x and y axes correspond to the first and the second largest components, respectively. The ratio of vertices with degree 3 contributed to 28\% in the first component while the average local transitivity contributes to 15\% of the second component.}
   \label{fig:featspca}
\end{figure}

The principal components of all graph measurements have been calculated (PCA) and the result is shown in Fig.~\ref{fig:featspca}. The obtained PCA projection accounts for 43\% of the overall variance, indicating that the projection onto two axes allows a reasonable representation of the original data.
The distribution of the cities in the PCA resulted with density decreasing from right to left, being more sparse at the latter case. At least four groups can be observed: one including five of the rightmost cities, namely Emdem, Bristol, Jena and Gelsenkirchen and Worms; one comprising Kassel, Wilhelmshaven and Liverpool; one comprising Santa Comba D'ao, Castelo de Paiva and Vila Nova de Cerveira; and the last, of Istres and Gaphautes-alpes. This behaviour considering the features analyzed, tends to suggest substantial similarity (homogeneiry) of the cities within each of these groups. In fact, the first group is mainly composed of German cities, the third is composed of three Portuguese cities, and the last is composed of two French cities. Paris resulted as the leftmost city, also being substantially separated from all other considered cities. It could be expected that cities that are near in the PCA present similar topological properties, being potentially related to higher gain.

\subsection{Distance reduction}

The average gain values obtained for the cities, considering all possible express way positions, are indicated in Figure~\ref{tab:gains}. Paris had the lowest gain in average shortest path length. This is likely due to the city having an efficient street network~\cite{costa2010efficiency}. The largest gain values are observed for the city of Santa Comba Dão. 

\begin{table}[]
    \centering
    \caption{Average gain considering the proposed approach.}
    \label{tab:gains}
    \begin{tabular}{@{}lr@{}}
        \toprule
        City & Average gain \\
        \midrule
        Paris & 0.002\\
        Liverpool & 0.006\\
        Kassel & 0.008\\
        Gelsenkirchen & 0.011\\
        Gaphautes-alpes & 0.013\\
        Heidelberg & 0.014\\
        Wilhelmshaven & 0.016\\
        Bristol & 0.017\\
        Istres & 0.024\\
        Emden & 0.025\\
        Jena & 0.027 \\
        Worms & 0.035 \\
        Castelo de Paiva & 0.038 \\
        Vila Nova de Cerveira & 0.053 \\
        Koblenz & 0.053 \\
        Estarreja & 0.058 \\
        Livorno & 0.090 \\
        Santa Comba Dão & 0.126 \\
        \bottomrule
    \end{tabular}
\end{table}

\begin{figure}[]
    \includegraphics[width=0.85\textwidth]{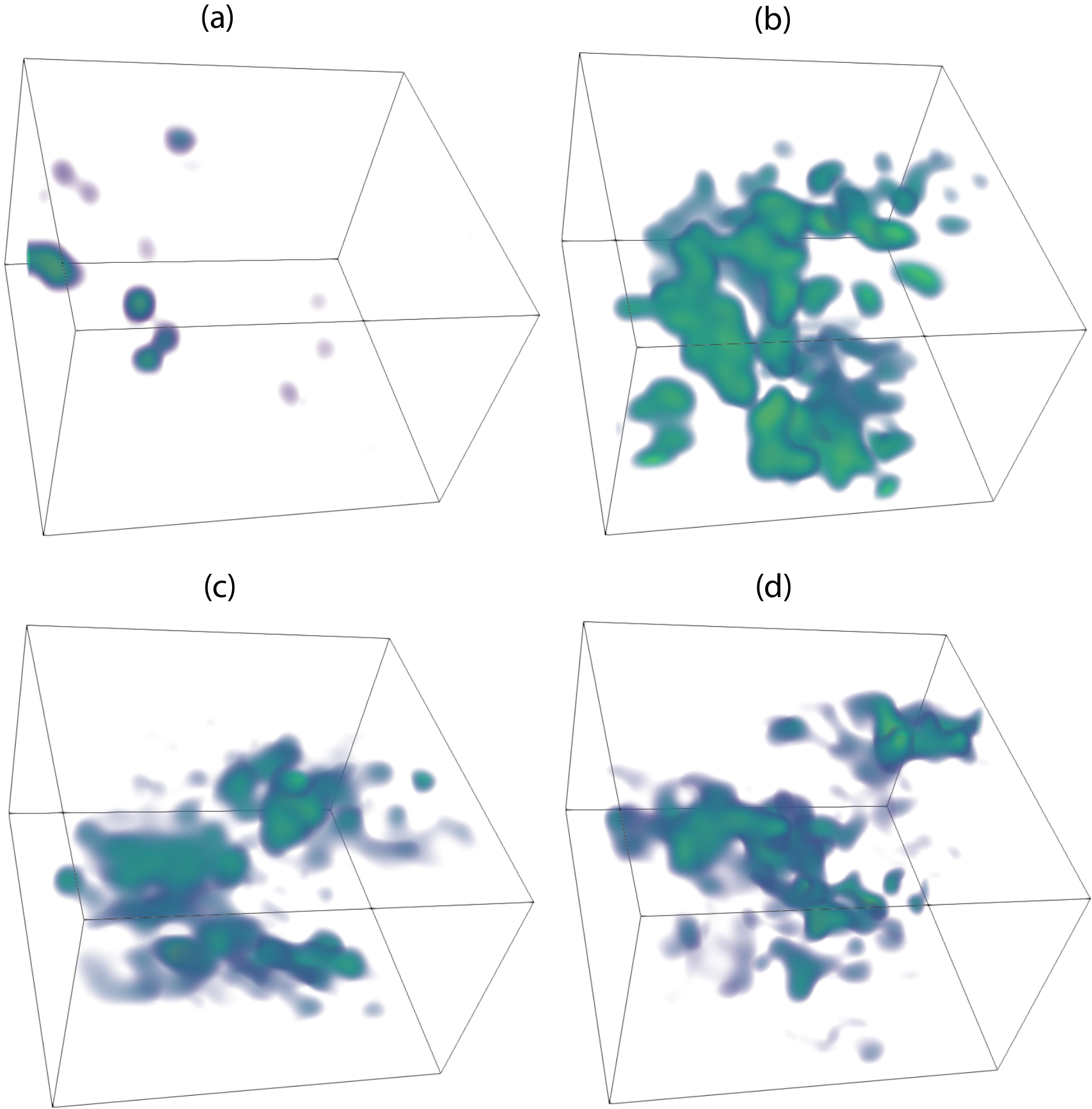}
   \caption{Visualization of volumes generated from the gains calculated for selected cities. Two axes of each volume correspond to the $(x,y)$ position of the express ways and the third axis is associated to the angles of the express ways. The figure shows the city of (a) Livorno, (b) Santa Comba Dão, (c) Paris and (d) Kassel.}
   \label{fig:gain_volumes}
\end{figure}

The gains calculated for all express ways where used for generating a 3D gain image for each city, as described in Section~\ref{subsec:gainanalysis}. Figure~\ref{fig:gain_volumes} shows examples of the generated images. Table~\ref{tab:gain_image_stats} shows the number of peaks and raggedness obtained for the images. A threshold of $t=0.001$ was used for calculating the raggedness. The influence of the number of peaks and the raggedness on the gain depend on how similar the gains are among the peaks. For instance, the presence of several peaks with similar gain values is desirable because the chances of finding a suitable solution is higher. On the other hand, if the gain values are very different among the peaks, it will be much more difficult to find a peak allowing a good solution. In cases characterized by a small number of peaks, it can be much more difficult to find, through a local procedure such as greedy search, the best express way placement solution. As for the effect of the raggedness on improving the gain, it could be expected that a more intricate gain distribution will typically make it harder to identify the global optimum.

The city of Livorno resulted in the smallest raggedness value. As can be seen in Figure~\ref{fig:gain_volumes}(a), the largest gains obtained for this city are mainly located in specific regions of the $R\times\Theta$ space. Interestingly, this city also resulted in the second largest average gain (as indicated in Table~\ref{tab:gains}. Thus, even though it is difficult to find proper positions for the express ways in this city, a great improvement can be obtained regarding the average shortest path length. Contrariwise, Santa Comba Dão has the largest raggedness. As can be seen in Figure~\ref{fig:gain_volumes}(b), the gain distribution for this city indeed has large complexity. Interestingly, Santa Comba Dão has many regions associated with large gains, but the variation of the gain inside each region is relatively smooth. Thus, the city does not contain many gain peaks. Actually, the number of peaks is similar between Santa Comba Dão and Livorno. 

Figure~\ref{fig:gain_volumes} also shows the gain volumes obtained for Paris (Figure~\ref{fig:gain_volumes}(c)) and Kassel (Figure~\ref{fig:gain_volumes}(d)). Both cities have a relatively complex gain distribution, which is reflected by their large raggedness value. The city of Kassel, Worms and Castelo de Paiva resulted in the largest number of peaks among the considered cities. Thus, these cities contain many different locations where the gain can be made locally maximal, which can hinder the identification of the global optimum.

Comparing Tables \ref{tab:gains} and \ref{tab:gain_image_stats}, we verify that the raggedness and the number of peaks are not strongly associated with the average gains of the cities. This is so because a single peak may be related to large gains, independently of the existence of other peaks. In a similar fashion, both smooth and complex gain changes can lead to significant increases of the gain. Still, the raggedness and the number of peaks are associated with the difficult in finding good express way placements. A single, large and smooth peak make it much easier for identifying the optimal point for reducing the average shortest path length of a city.

\begin{table}[]
    \centering
    \caption{Raggedness and number of peaks calculated for the gains obtained for the analysed cities.}
    \label{tab:gain_image_stats}
    \begin{tabular}{lrr}
    \toprule
    {} & Raggedness & Number of peaks \\
    \midrule
    Livorno            &      0.74 &         8 \\
    Istres             &      1.49 &        10 \\
    Gaphautes-alpes    &      1.56 &         7 \\
    Koblenz            &      1.56 &         6 \\
    Bristol            &      1.57 &         9 \\
    Gelsenkirchen      &      1.64 &         9 \\
    Worms              &      1.64 &        11 \\
    Heidelberg         &      1.67 &         8 \\
    Liverpool          &      1.70 &         9 \\
    Emden              &      1.71 &         8 \\
    Wilhelmshaven      &      1.73 &         9 \\
    Kassel             &      1.74 &        11 \\
    Paris              &      1.76 &         7 \\
    Jena               &      1.77 &         8 \\
    Vila Nova de Cerveira &      1.81 &         5 \\
    Castelo de Paiva     &      1.82 &        11 \\
    Estarreja          &      1.82 &         8 \\
    Santa Comba Dão      &      1.88 &         7 \\
    \bottomrule
    \end{tabular}
\end{table}

\section{Correlation between gain and topological features}

\begin{table}[]
	\caption{Pearson correlation between the graph features and the average gain obtained.}
	\label{tab:corrtopgain}
	\centering
    \begin{tabular}{@{}lr@{}}
        \toprule
        Feature & Correlation \\
        \midrule
	acstd & -0.483 \\
        acavg & -0.267 \\
        degr5 & -0.146 \\
        degr4 & -0.099 \\
        destd & 0.003 \\
        degr3 & 0.029 \\
        deavg & 0.084 \\
        vpstd & 0.292 \\
        eostd & 0.367 \\
        trstd & 0.486 \\
        travg & 0.496 \\
        lacun & 0.596 \\
        \bottomrule
    \end{tabular}

\end{table}

The correlation of each measurement and the average gain was calculated and the results are presented in Table~\ref{tab:corrtopgain}. In order to better understand and try to explain the obtained average gains,
we focus on the topological and geometrical features that led to the highest
correlations with the gain, which corresponds to lacun and transmean.
These two cases are discussed next.

\begin{figure}[]
    \begin{subfigure}[b]{0.47\textwidth}
   \includegraphics[width=\textwidth]{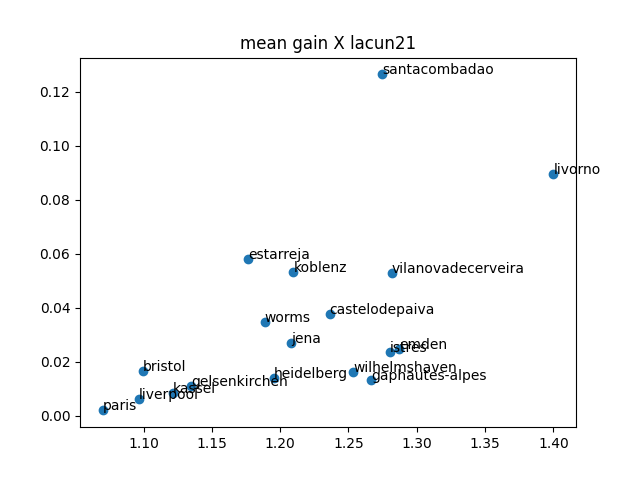}
       \centering
  \caption{}
    \end{subfigure} \quad
    \caption{Average path length gain and city features. Relationship between the average path length gain (y axis) with the lacunarity.}
    
    \label{fig:corranglacun}
\end{figure}

Figure~\ref{fig:corranglacun}(a) shows the scatterplot obtained considering 
the lacunarity represented in the horizontal axis, while the distance gain
is shown along the vertical axis.  At the lower left corner of this plot, we
find Paris, which is characterized by the smallest values of both lacunarity and 
distance gain.  As low lacunarity values tend to indicate a more ordered geometry
of city blocks, we can conclude that the gain in this case is the smallest because
Paris is already spatially very uniform.  A similar conclusion can be reached
regarding the neighboring cities (in the scatterplot) of Bristol, Liverpool and
Kassel.  As the lacunarity values increase (moving rightward along the horizontal
axis), the gain values tend to become more disperse (for a fixed lacunarity),
suggesting that a less definite relationship can be found for the
less spatially uniform cities.  For instance, for lacunarities around $1.30$,
we have Istres presenting gain of 0.02, Villa nova with gain 0.05 and Santa Comba Dão
with a distance gain of 0.13.  The city of Livorno yielded the largest lacunarity
value of 1.40 and a relatively large respective gain, 0.09 further corroborating the
tendency of higher gains being observed for less spatially uniform cities.

\section{Concluding remarks}

Several structures and systems from the natural as well as human worlds are optimized to
perform specific functionalities.  For instance, the structure and organization of cities 
need to cater for expedite mobility and access to the most important resources such
as hospitals, fire stations, schools, etc.  These displacements depend critically on the
shortest path between any two points in a city.  It therefore constitutes
a particularly important endeavor trying to improve the overall accessibility and mobility
in any given city by incorporating new express ways, such as avenues, bridges and tunnels.

In the present work, we developed a systematic approach aimed at obtaining complete energy
surfaces concerning the shortest path gain implied by the incorporation of a express way
between several combinations of two points in a given city.  While one of the extremities
is fixed along a grid imposed on the city, the other point is placed at a fixed distance
of 0.25 times the city diameter, but according to several uniformly spaced angles.  A complete 3D surface of shortest path gains can therefore be obtained that provides a particularly
comprehensive and interesting characterization of the potential of each city to shortest
path improvements through the incorporation of an express way.  

The energy surfaces obtained respectively to 18 European cities with similar areas resulted
very distinct one another, reflecting quite different potentials for improvement intrinsic
to each of the considered cities.  These surfaces are typically characterized by possessing
several extreme points, which are distributed in more or less homogeneous manner along
energy surfaces that can be more or less intricate.  These
specific distributions underlie not only the potential for improvement, but also how
difficult it may be to build the express ways in the optimal positions. For instance, for cities possessing peaks with low gain values, the average shortest path length cannot be significantly improved, as in the case of the city of Paris. If many peaks with heterogeneous gain values are observed, it can be difficult to identify proper express ways placements for large gains, specially in the cases of complex gain surfaces, since local optima can hinder the identification of the global optimum. When only a few peaks with heterogeneous values are present in the city, it may also be difficult to identify the correct positions for express ways placement. On the other hand, a large number of high valued peaks distributed uniformly along the optimization space
implies improvements of shortest distance for almost any incorporated express
way. 

While the potential for gain improvement was correlated with the lacunarity and transitivity, the correlation between the measurements associated with the gain surface (number of peaks and raggedness) resulted in low correlation with topological and geometrical measurements.

Future analyses may consider additional cities and measurements, as well as a finer grid for express way placement. Also, further investigations are necessary for better understanding the relationship between the gain surface and the topological and geometrical measurements of the cities.

\section*{Aknowledgments}
EKT thanks FAPESP (grant 2019/01077-3). CHC thanks FAPESP (grant no. 18/09125-4 and 2021/12354-8) for financial support. LFC thanks CNPq (grant no. 307085/2018-0) for sponsorship.  The authors also acknowledge FAPESP grant 15/22308-2 and CAPES.

\bibliographystyle{plain}
\bibliography{refs}

\begin{thebibliography}{10}

\bibitem{ahuja1995applications}
Ravindra~K Ahuja, Thomas~L Magnanti, James~B Orlin, and MR~Reddy.
\newblock Applications of network optimization.
\newblock {\em Handbooks in Operations Research and Management Science},
  7:1--83, 1995.

\bibitem{allain1991characterizing}
Cloitre Allain and M~Cloitre.
\newblock Characterizing the lacunarity of random and deterministic fractal
  sets.
\newblock {\em Physical review A}, 44(6):3552, 1991.

\bibitem{barthelemy2011spatial}
Marc Barth{\'e}lemy.
\newblock Spatial networks.
\newblock {\em Physics Reports}, 499(1-3):1--101, 2011.

\bibitem{buhl2006topological}
Jerome Buhl, Jacques Gautrais, N~Reeves, Ricard~V Sol{\'e}, Sergi Valverde,
  Pascale Kuntz, and Guy Theraulaz.
\newblock Topological patterns in street networks of self-organized urban
  settlements.
\newblock {\em The European Physical Journal B-Condensed Matter and Complex
  Systems}, 49(4):513--522, 2006.

\bibitem{costa2007characterization}
L~da~F Costa, Francisco~A Rodrigues, Gonzalo Travieso, and Paulino~Ribeiro
  Villas~Boas.
\newblock Characterization of complex networks: A survey of measurements.
\newblock {\em Advances in physics}, 56(1):167--242, 2007.

\bibitem{costa2010efficiency}
L~da~F Costa, BAN Traven{\c{c}}olo, MP~Viana, and E~Strano.
\newblock On the efficiency of transportation systems in large cities.
\newblock {\em EPL (Europhysics Letters)}, 91(1):18003, 2010.

\bibitem{costa2021comparing}
Luciano da~F Costa.
\newblock Comparing cross correlation-based similarities.
\newblock {\em arXiv preprint arXiv:2111.08513}, 2021.

\bibitem{costa2021further}
Luciano da~F Costa.
\newblock {Further generalizations of the Jaccard index}.
\newblock {\em arXiv preprint arXiv:2110.09619}, 2021.

\bibitem{da2010shape}
Luciano da~Fontoura~Costa and Roberto~Marcondes Cesar~Jr.
\newblock {\em Shape analysis and classification: theory and practice}.
\newblock CRC press, 2010.

\bibitem{demaine2010minimizing}
Erik~D Demaine and Morteza Zadimoghaddam.
\newblock Minimizing the diameter of a network using shortcut edges.
\newblock In {\em Scandinavian Workshop on Algorithm Theory}, pages 420--431.
  Springer, 2010.

\bibitem{frati2015augmenting}
Fabrizio Frati, Serge Gaspers, Joachim Gudmundsson, and Luke Mathieson.
\newblock Augmenting graphs to minimize the diameter.
\newblock {\em Algorithmica}, 72(4):995--1010, 2015.

\bibitem{hildebrandt1996parsimonious}
Stefan Hildebrandt and Anthony Tromba.
\newblock {\em The parsimonious universe: shape and form in the natural world}.
\newblock Springer Science \& Business Media, 1996.

\bibitem{luo2008computing}
Jun Luo and Christian Wulff-Nilsen.
\newblock Computing best and worst shortcuts of graphs embedded in metric
  spaces.
\newblock In {\em International Symposium on Algorithms and Computation}, pages
  764--775. Springer, 2008.

\bibitem{meyerson2009minimizing}
Adam Meyerson and Brian Tagiku.
\newblock Minimizing average shortest path distances via shortcut edge
  addition.
\newblock In {\em Approximation, Randomization, and Combinatorial Optimization.
  Algorithms and Techniques}, pages 272--285. Springer, 2009.

\bibitem{newman2001scientific}
Mark~EJ Newman.
\newblock Scientific collaboration networks. i. network construction and
  fundamental results.
\newblock {\em Physical review E}, 64(1):016131, 2001.

\bibitem{pinkney2019napoleon}
David~H Pinkney.
\newblock {\em Napoleon III and the Rebuilding of Paris}.
\newblock Princeton University Press, 2019.

\bibitem{rijkersaugmenting}
Stefan Rijkers.
\newblock Augmenting geometric networks computing the best shortcut.
\newblock Master's thesis, Technische Universiteit Eindhoven, Department of
  Mathematics and Computer Science, 2010.

\bibitem{rodrigues2005self}
Erbe~P Rodrigues, Marconi~S Barbosa, and Luciano da~F Costa.
\newblock Self-referred approach to lacunarity.
\newblock {\em Physical Review E}, 72(1):016707, 2005.

\bibitem{sewell1987maximum}
Michael~J Sewell et~al.
\newblock {\em Maximum and minimum principles: a unified approach with
  applications}, volume~1.
\newblock CUP Archive, 1987.

\bibitem{travenccolo2008accessibility}
Bruno Augusto~Nassif Traven{\c{c}}olo and L~da~F Costa.
\newblock Accessibility in complex networks.
\newblock {\em Physics Letters A}, 373(1):89--95, 2008.

\bibitem{troutman2012variational}
John~L Troutman.
\newblock {\em Variational calculus and optimal control: optimization with
  elementary convexity}.
\newblock Springer Science \& Business Media, 2012.

\bibitem{zhan1998shortest}
F~Benjamin Zhan and Charles~E Noon.
\newblock Shortest path algorithms: an evaluation using real road networks.
\newblock {\em Transportation science}, 32(1):65--73, 1998.

\end{thebibliography}
\end{document}